\documentstyle[prl,aps,epsf,multicol]{revtex}
\begin{document}
\date{\today}
\title {Spin dependent scattering of a domain-wall of controlled 
size.}
\author{J.-E. Wegrowe \cite{email}, A. Comment, Y. Jaccard, J.-Ph. 
Ansermet}
\address{Institut de Physique Exp\'erimentale, Ecole Polytechnique 
F\'ed\'erale de Lausanne, CH - 1015 Lausanne, Switzerland.}
\author{N. M. Dempsey, J-P. Nozi\`eres}
\address{Laboratoire de magn\'etisme Louis N\'eel, CNRS, BP166X 38042 
Grenoble, France.}
\maketitle

\begin{abstract}

Magnetoresistance measurements in the CPP geometry have been performed on 
single electrodeposited Co nanowires exchange biased on one side by a 
sputtered amorphous GdCo$_{1.6}$ layer.  This geometry allows the stabilization 
of a single domain wall in the Co wire, the thickness of which can be 
controlled by an external magnetic field.  Comparing magnetization, 
resistivity, and magnetoresistance studies of single Co nanowires, of 
GdCo$_{1.6}$ layers, and of the coupled system, gives evidence for an additional 
contribution to the magnetoresistance when the domain wall is compressed by 
a magnetic field.  This contribution is interpreted as the spin dependent 
scattering within the domain wall when the wall thickness becomes smaller 
than the spin diffusion length.

\end{abstract}


\begin{multicols}{2}
\narrowtext

\section{Introduction}

Spin dependent scattering studies emerged with the first realizations 
of magnetic nanostructures, and gave rise to the discovery on spin 
injection \cite{1}, Giant magnetoresistance (GMR) \cite{2,3} and 
tunneling magnetoresistance (TMR) \cite{39}.  In 
ferromagnets, Spin polarization of the current is due to spin-flip 
scattering which favors the orientation of the conduction electron 
spins in the direction of the magnetization.  When a change of 
magnetization occurs at the nanoscopic scale, the spins of conduction 
electrons relax from the initial to the final direction of the 
magnetization, and this relaxation leads to an supplementary 
resistance \cite{1,4,5}.  However, this relaxation takes place only in 
the case of an abrupt spatial change of the magnetization, namely a 
length smaller or equal to the spin diffusion length $l_{sf}$ (about 
$20 \, nm$ in Co measured with Co/Cu/Co multilayers) \cite{38,6} .  
In the case of a smooth magnetic 
change, the polarization axis of the conduction electrons follows the 
local magnetic field adiabatically \cite{7,8} and there is no 
relaxation.
 
 The spin relaxation in artificial magnetic multilayers with 
 anti-parallel coupling has been studied extensively both at the 
 experimental and theoretical level due to the large magnetoresistance 
 observed (the so-called GMR) \cite{9}.  However, in the case of non 
 uniformly magnetized layers, or domain-walls (the so-called 
 Domain-wall Scattering problem: DWS), the relaxation has not been 
 evidenced clearly.  Different theoretical models were proposed and 
 are currently under development \cite{8,10,11,12}.  Simply speaking, 
 two regimes have to be considered depending on the domain-wall 
 thickness $\delta$.  On one hand, for domain-wall thickness larger 
 than a characteristic length (let say $\delta > l_{sf}$), the spin 
 polarization axis rotates adiabatically and hence no supplementary 
 resistance is expected.  On the other hand, as $\delta$ approaches 
 this characteristic length, a progressive enhancement of the relaxation 
 occurs, resulting in an additional magnetoresistance contribution.

  Several attempts were recently made to measure the DWS 
  \cite{14,15,16,17,18,37}.  The main difficulty of these studies 
  appears to be due to the importance of the bulk anisotropic 
  magnetoresistance (AMR) which is due to the anisotropy of the 
  spin-orbit scattering, with respect to the direction of the current 
  \cite{19,20}.  In order to correct the AMR contribution, the 
  micromagnetic configurations have to be determined with great 
  precision. However, the domain-walls are present at low field, where 
  other non uniform spin configurations usually coexists (like flux 
  closures).  Furthermore, at low external field the domain-wall 
  thickness is defined by the local magnetostatic and anisotropy 
  field, and can not be controlled easily.

The aim of the present study is to measure the DWS in a nanoscopic 
domain-wall the thickness of which can be controlled by a strong 
external magnetic field.  The measurement is performed in the Current 
Perpendicular to the domain-wall (CPW) geometry according to a 
procedure already developed for measuring single nanowires \cite{23}.  
A "Zeeman Domain-wall" was constructed by employing a ferrimagnetic 
amorphous GdCo$_{1.6}$ alloy as a pinning layer on one side of a Co 
nanowire.  In amorphous GdCo alloys at low temperature, the magnetic 
moments of Gd (about $7 \, \mu _B$) and of Co (about $2 \, \mu _B$) 
are aligned ferrimagnetically.  When the Gd contribution predominates 
(which is the case here), the Co moments are always aligned 
anti-parallel with respect to an external magnetic field, as long as 
the amplitude of the field is smaller than the Gd-Co atomic exchange 
field (see Fig.1(a)).  Consequently, near the 
GdCo$_{1.6}$/Co interface, where the Co-Co exchange is dominant, a 
stable 180° domain-wall, which is compressed by the external field, 
exists in the Co sublattice.  Details of the micromagnetic 
configuration of such bilayers can be found in \cite{21,22}.

The structure of this paper is the following. The 
samples preparation (Section I) is first described. 
The magnetization characterization 
(Section II), the temperature dependence of the resistivity (Part III), 
and the magnetoresistance (Section IV) of (A) Co nanowires, (B) 
GdCo$_{1.6}$ thin films, and (C) hybrid Co/GdCo$_{1.6}$ samples are 
then reported. Analyse and discussion about the action of the 
interface and domain-wall scattering follow in Section V. The 
domain-wall thickness $\delta (H,T)$ is evaluated by using a simple 
model. A scaling plot $R(\delta (H,T))$ yields the MR as a 
function of the domain-wall thickness.

\section{Samples preparation}

Nanocrystallized Co nanowires are obtained by electrodeposition in 
track etched membrane templates \cite{24}.  Polyester membranes with about $6 
\cdot 10^8$ pores/$cm^2$ were used.  The wires are about 6000 nm in 
length and 80 nm in diameter.  An in-situ technique, inside the 
electrolytic bath, allows a sub-micron contact to be made, in 
order to measure one single nanowire \cite{23}. The detailed 
procedure is as follows:

(1) One surface (called in the following  S2) of the membrane is recovered by a 
$200 \, nm$ thick sputtered gold layer. This surface plays the role 
of working electrode during the Co electrodeposition. The other 
surface (surface S1 in Fig.1(b)) is open to the Co electrolytic bath. 
The pores are filled with Co up to about a third of the pores length. 

(2) After cleaning the surface S2 with potassium cyanide (KCN) in 
order to dissolve totally the gold electrode, a film of $220 \, 
nm$ amorphous GdCo is sputtered at room temperature from a 
GdCo$_{2}$ target. The GdCo layer is hence in contact with the bottom of the Co 
wires. A $20 \, nm$ Mo protection layer is deposited on the GdCo 
film to avoid oxidation. 

(3) A gold film of $50 \, nm$ is sputtered on the surface S1 in order 
to control the potential between the two surfaces of the membrane 
during electrodeposition. This thin layer does not obstruct the 
pores. The surface S2 play the role of working electrode again. The 
surface S1 and the top of the Co wires are cleaned by "ex-situ" reduction in 
a NaCl electrolytic bath in order to remove the skin of the Co oxide. 
(4) The rest of the pores are filled with Co while controlling the 
voltage between the surfaces S1 and S2. A single contact is obtained by 
a feed-back loop which stops the deposition as soon as the voltage 
reaches zero (i.e. when the first wire connects to the surface S1). 

\vspace{-0.7cm}
\begin{figure}[tbp]
\epsfxsize=7cm
$$
\epsfbox{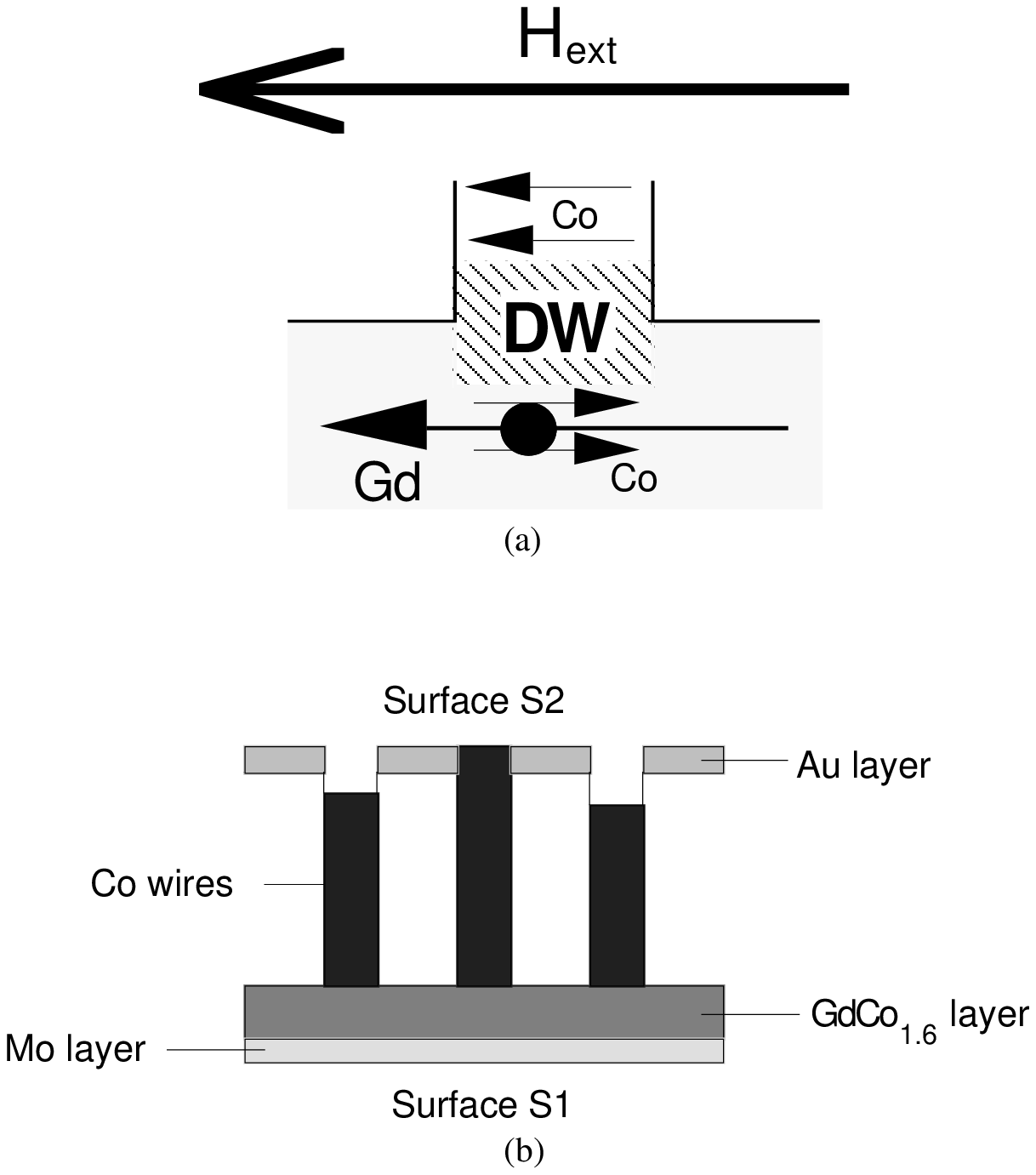}
$${fig1.eps}
    \caption{(a) Schematic diagram of an interface between a Co wire and the 
GdCo$_{1.6}$ layer. The presence of a $180^o$ domain wall is 
evidenced by the hatched zone. (b) General structure of the hybrid 
GdCo/Co system.}
    \label{fig:fig1}
\end{figure}€

The ferromagnetic wires obtained by this method have been 
characterized by various techniques, including NMR spectroscopy and X 
ray diffraction \cite{25}.  It was shown that the Co is 
nano-crystallized with a broad size distribution of crystallites sizes 
and with a mixture of fcc and hcp phases.  The GdCo$_{x}$ layer 
stochiometry was estimated, by electron dispersive x-ray (EDX) 
analysis, to be GdCo$_{1.6}$.

\section{Magnetic characterization}

\subsection{Co nanowires}

The magnetic characterization of the Co wires has been performed on 
a membrane (containing about $10^7$ nanowires),
with the field perpendicular to the wires axes (Fig.2). It has been 
shown that the magnetocrystalline anisotropy direction of the 
crystallites are distributed  
perpendicular to the wire axis, due to strains and magnetostriction 
during the growth within the pores 
\cite{26}. The remanent state is not uniformly magnetized along the 
wire axis, and a domain-wall or vortex is present
\cite{27}. The spin configuration of this state is very sensitive to 
the nanocrystalline configuration.

\vspace{-0.7cm}
\begin{figure}[tbp]
\epsfxsize=7cm
$$
\epsfbox{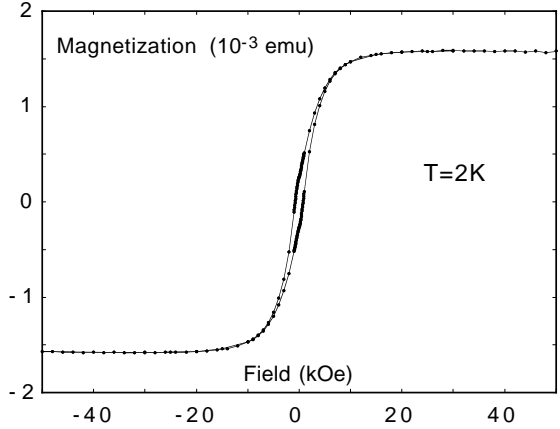}
$$
    \caption{Hysteresis loop of a membrane with about $10^7$ 
Co nanowires, measured at $2 \, K$, with the applied field perpendicular 
to the wire.}
    \label{fig:fig2}
\end{figure}€

\subsection{GdCo$_{1.6}$ thin films}

The temperature dependence of the GdCo$_{1.6}$ has been reported in 
reference
 \cite{28}. The magnetic moment is 
almost zero at room temperature and increases with decreasing
temperature, according to the curve of Fig.3(a), measured under a 5T 
external field. The 
shape of the magnetic hysteresis loop at $2 \, K$ show that the 
magnetization at $5 \, T$ is not the saturation magnetization, as 
expected for ferrimagnetic materials (Fig.3(b)).

\vspace{-0.7cm}
\begin{figure}[tbp]
  \epsfxsize=7cm
    \epsfbox{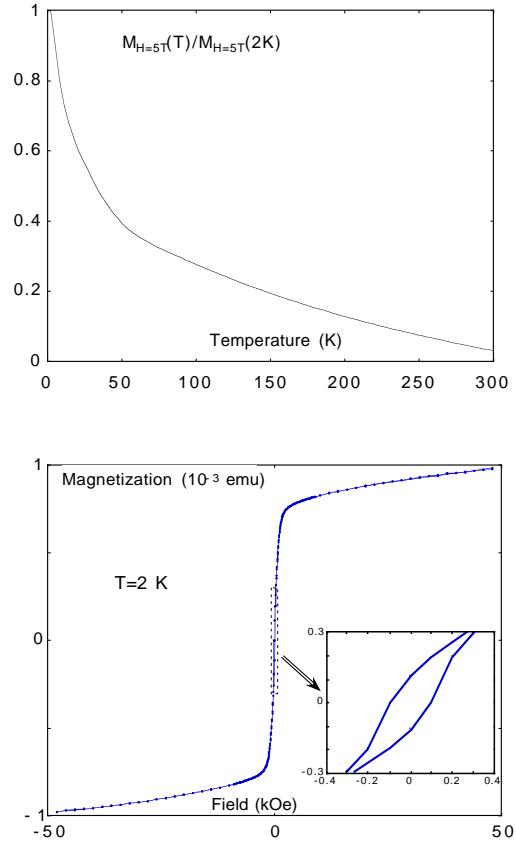}
    \caption{(a) Temperature dependence of the magnetization of a
GdCo$_{1.6}$ layer measured at $H = 5 \, T$ in the 
plane of the layer. (b) hysteresis loop of the 
GdCo$_{1.6}$ layer measured at $2 \, K$ in the plane of the layer.}
    \label{fig:fig3}
\end{figure}€

\subsection{GdCo$_{1.6}$/Co hybrid structures}

 At high temperature, GdCo$_{1.6}$ moment is too weak to provide a 
 measurable exchange bias.  The hybrid structure therefore behaves 
 like the Co nanowires.  The magnetic hysteresis loop of the hybrid 
 structure at $2 \, K$ is shown in fig.4, with to the sum of the 
 individual contribution of the Co nanowires and the GdCo$_{1.6}$ 
 layer.  The difference between the two curves of fig.4 (hatched area 
 in Fig4) is due to the effect of the exchange biasing at the 
 interface, leading to the creation of a large domain-wall at low 
 field.  The magnetic curve of the hybrid structure can hence be 
 described, in a first approximation, by the relation

\begin{eqnarray}
M_{GdCo_{1.6}/Co}(H)
&=& M_{GdCo_{1.6}}(H)+\\
& & + [1 - \delta (H)/L] \cdot M_{Co}(H)\nonumber
\end{eqnarray}

\noindent where L is the total length of the Co wire i.e. $L\,=\, 6000 \, 
nm$. Above $3 \, T$, the two loops are superimposed. The contribution 
of the domain-wall $\delta (H)/L$ is hence 
negligible for $H > 3 \, T$: the contribution of the DW to the AMR 
should hence be also negligible at high field.
\vspace{0.7cm}
\begin{figure}[tbp]
    \epsfxsize=8cm
    \epsfbox{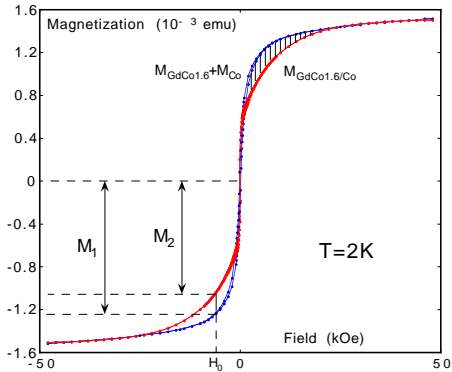}
    \caption{$2 \, K$ hysteresis loop of the GdCo$_{1.6}$ /Co hybrid structure 
superimposed with the sum of the individual hysteresis loops of the 
Co wire and the GdCo$_{1.6}$ layer. The difference 
$M_1-M_2$ (hatched area) is the contribution of the low field domain wall.}
    \label{fig:fig4}
\end{figure}€

\section{Electrical characterization}

\subsection{Co nanowires}

Resistivity measurements have been performed on Co samples. The 
general temperature profile is plotted in Fig.5. The value of the 
resistance,($50 \, \Omega$), indicates that two or three wires are 
contacted in parallel in this sample. The general profile is in 
accordance with the behavior expected for bulk Co. The 
contribution due to phonon scattering is linear and the curvature is 
due to the spin-disorder resistivity, observable only at high 
temperature \cite{29}. The residual resistivity $R_0$ is rather sensitive to the Co 
nanocrystalline structure and the quality of the micro-contact, which 
are seldom reproducible from one wire to an other. 

\vspace{-0.7cm}
\begin{figure}[tbp]
    \epsfxsize=7cm
  \epsfbox{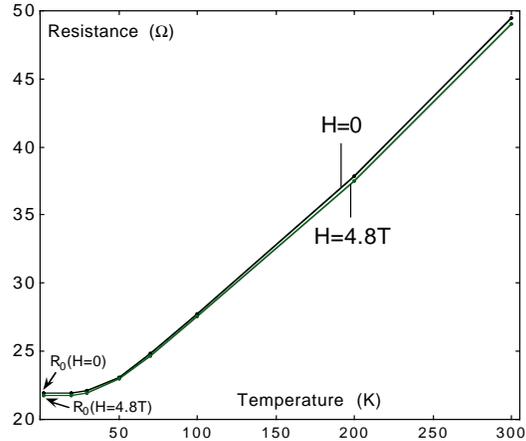}
    \caption{Temperature dependence of the resistance of the Co nanowire 
at $H=0$ and $H=4.8 \, T$.}
    \label{fig:fig5}
\end{figure}€

\subsection{GdCo$_{1.6}$ thin films}

The transport properties are measured in the current-in-plan geometry 
(CIP) with the standard 4 probe technique.  The resistance of the thin 
films in this configuration is of the order of $5 \, \Omega$ at room 
temperature.  The low temperature profile (Fig.  6), well reproducible, 
shows a typical feature, characteristic of the "structural Kondo 
effect" \cite{30,31}: this profile does not depend on the external 
magnetic field (see section IV), and hence it is not caused by spin 
dependent scattering.  In contrast to the magnetic Kondo effect, where 
the negative thermal coefficient of the resistivity is due to magnetic 
impurities dilute in 
non-magnetic metals, the "structural Kondo effect" is due (in an 
oversimplified picture \cite{31}) to an other degree of freedom 
present in disordered systems. The temperature dependence 
of the resistivity below about $20 \, K$ follows the form $\rho (T) = 
-\alpha \cdot ln(k_B^2 \, T^2 + \Delta ^2) + \beta$ where $\alpha$ is 
a positive coefficient, $k_B$ the Boltzmann constant and $\Delta$, 
$\beta$ constants.  This behaviour is in agreement with previous 
studies on Gd-Co amorphous films \cite{32}.

\vspace{0.7cm}
\begin{figure}[tbp]
    \epsfxsize=7cm
    \epsfbox{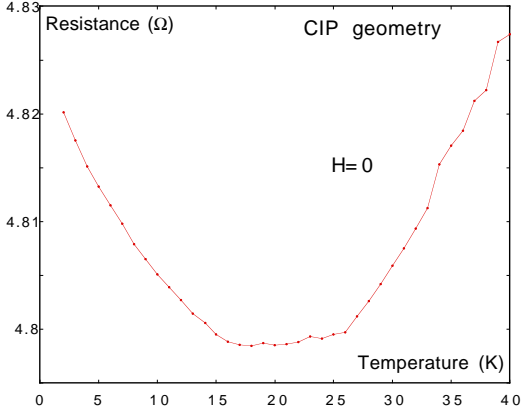}
    \caption{Temperature dependence of the resistance of a 
GdCo$_{1.6}$  layer at $H=0$, measured in CIP geometry.}
    \label{fig:fig6}
\end{figure}€

\subsection{GdCo$_{1.6}$/Co hybrid structures}

The resistance of the hybrid structure is about $180 \, \Omega$ at 
room temperature. The contributions of the Co wire and of the 
GdCo$_{1.6}$ film are observed in the temperature dependence of the 
resistance (Fig.7). From $300 \, K$ down to $50 \, K$, the profile 
follows the usual bulk Co curve, while below $T_{min} = 20 \, K$ the structural Kondo 
effect dominates (inset of Fig.7) at zero magnetic field and 
$4.8 \, T$ magnetic field.
\vspace{-0.7cm}
\begin{figure}[tbp]
   \epsfxsize=7cm
    \epsfbox{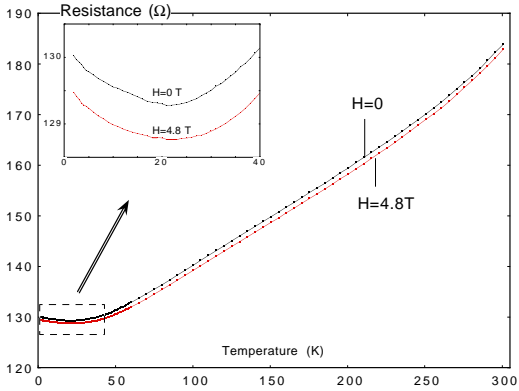}
    \caption{Temperature dependence of the resistance of the hybrid 
structure at $H=0$ and $H=4.8 \, T$. Inset: low temperature zoom.}
    \label{fig:fig7}
\end{figure}€

\section{Magnetoresistance}

\subsection{Co nanowires}

The magnetoresistive hysteresis loops of isolated Co wires at 
different temperatures are shown in Fig.8.  The magnetoresistance is 
normalized with respect to the zero field resistance $R(H=0)$ at each 
temperature.   The room temperature
magnetoresistance has been studied in 
detail elsewhere \cite{26}. The behavior observed is typical of anisotropic 
magnetoresistance (AMR) in which the resistance increases as the 
magnetization is aligned with the current. If $\phi $ is the angle 
between the current and the magnetization M(H), the magnetoresistance 
hysteresis loop is linked to the magnetic hystersis loop in the wire by the 
relation $\Delta R(H) = 
R_{0}+ \Delta R_{max}\,cos^{2}(\phi (H))$.  The jumps of the 
magnetization show the unstable states corresponding to the nucleation 
and the annihilation of domain-walls (Fig.8).  
\vspace{-0.7cm}
\begin{figure}[tbp]
    \epsfxsize=7cm
    \epsfbox{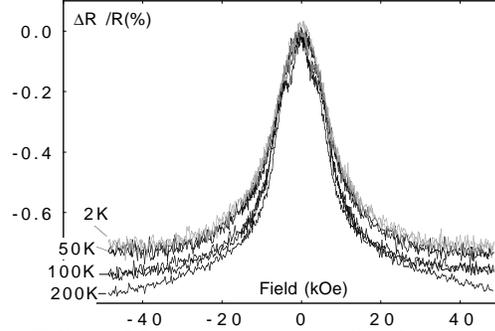}
    \caption{Magnetoresistance of a Co wire, normalized at 
$R(H=0)=50 \, \Omega$.}
    \label{fig:fig8}
\end{figure}€

On the other hand, the high field 
magnetoresistance shows an important temperature dependence.
As shown in Fig.9, the quantity $\Delta ^{4.8} R = 
R(H = 0) - R(H = 4.8 \, T)$ decreases linearly from $300 \, K$ 
($\Delta ^{4.8} R \approx 0.4 \Omega$) down to about $50 \, K$ 
($\Delta ^{4.8} R \approx 0.17 \Omega$).  This MR contribution is 
opposite to the Lorentz magnetoresistance.  It is characterized by the 
fact that the increase of the field reduces the resistivity in the 
same way that the decrease of the temperature does.  This behaviour 
may be explained by the spin-disorder, which contribution can be seen 
in the temperature dependence of the resistivity (Fig5).  The 
magnetoresistance due to this contribution is poorly understood.  
However, below $60 \, K$, the effect disappears and $\Delta ^{4.8} 
R(T)$ is roughly constant.  In the following, we will focus on the low 
temperature region only: $T < 60 \, K$.

\vspace{-0.7cm}
\begin{figure}[tbp]
    \epsfxsize=7cm
    \epsfbox{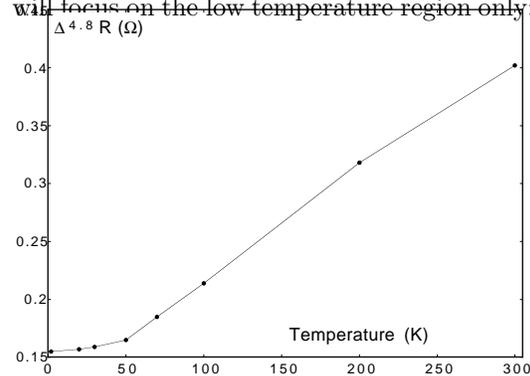}
    \caption{Temperature dependence of the anisotropic Co 
magnetoresistance $\Delta ^{4.8} R=R(H=0)-R(H=4.8 \, T$).}
    \label{fig:fig9}
\end{figure}€

\subsection{GdCo$_{1.6}$ thin films}

The low temperature magnetoresistance of GdCo thin films is reported 
in figure 10a and 10b for $H \perp I$ and $H // I$, respectively.  
Note that the asymmetry observed in Fig.10(a) is not relevant, as it 
is due to Hall currents in the 4 probe geometry.  The magnitude of 
$\Delta ^{4.8} R$ is quite similar and very small in the both configurations.
  The two magnetoresistance curves at 60K and 2K are approximately 
  superimposed, and only a 
weak AMR contribution can be observed (Fig.10(a)).  This is in 
accordance with the interpretation of section III(B) in terms of 
"structural Kondo effect" : there is no thermal dependence of this 
magnetoresistance. 

\vspace{-0.7cm}
\begin{figure}[tbp]
\epsfxsize=7cm
\epsfbox{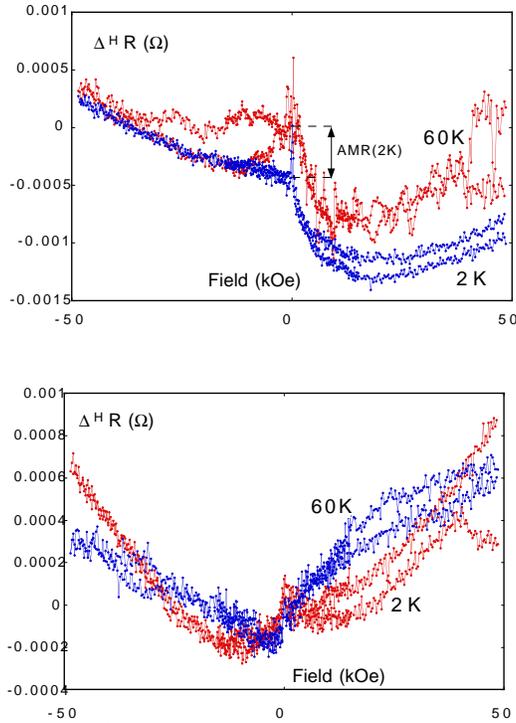}	
\caption{CIP magnetoresistance of the GdCo$_{1.6}$ layer with (a) the
applied field perpendicular to the current and (b) the 
applied field parallel to the current.}
\label{fig:fig10}
\end{figure}

\subsection{GdCo$_{1.6}$/Co hybrid structures}

 The magnetoresistance hysteresis loop is shown in Fig.11.  Due to the 
 exchange biasing, the switching fields are lowered 
 (Fig.11(b)) with respect to the Co switching fields, and
 a temperature variations of the remanent states can 
 be observed.  These effects are due to the temperature dependence 
 of the excahnge biasing at low field. The resulting 
 magnetic configurations at the interface are very difficult to 
describe, and the contribution of the 
 DWS, if any, cannot be extracted from the AMR at low field.

On the other hand, at high field ($H > 3T$), it as been shown that the contribution of 
the DW to the magnetization (Section II(C)), and hence to the AMR, is negligible. 
Since the (CIP) magnetoresistance of the GdCo$_{1.6}$ thin film is of the 
order of $10^{-3} \,\%$ of the total resistance (Section above), the 
magnetoresistance of the hybrid structures is expected to be identical 
to that of isolated Co wires at high field.

\vspace{-0.7cm}
\begin{figure}[tbp]
\epsfxsize=7cm
 \epsfbox{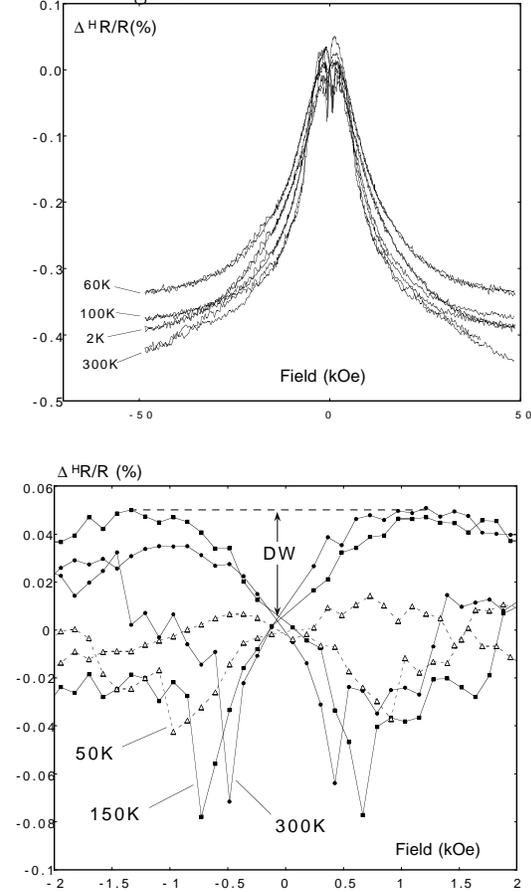}	
\caption{(a) Magnetoresistance of the GdCo$_{1.6}$/Co hybrid structure,
 normalized at $R(H=0)=180 \, \Omega$. (b) low field zoom.}
\label{fig:fig11}
\end{figure}

However, the high field magnetoresistance $\Delta ^HR/R$ behavior is 
not as expected (Fig.11(a)).
If the decrease of the AMR ratio, from $300 \, K$ down to 
$60 \, K$ is in accordance with the observation on the Co 
samples, the minimum of the high field magnetoresistance at about 
$T=60 \, K$ is surprising. This
temperature dependence is reploted in terms of magnetoresistance 
$\Delta ^HR(T)$ (Fig.12) for different applied fields.

\vspace{-0.7cm}
\begin{figure}[tbp]
\epsfxsize=7cm
\epsfbox{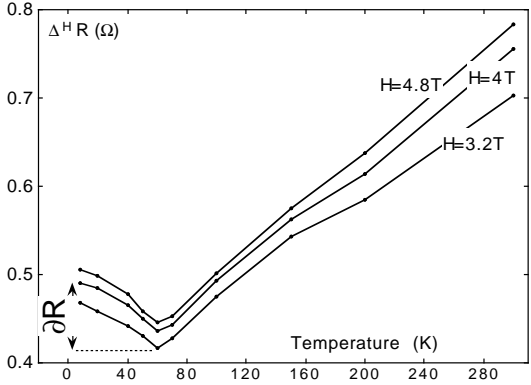}	
\caption{Temperature dependence of the GdCo$_{1.6}$/Co hybrid 
structure magnetoresistance $\Delta ^H R=R(H=0)-R(H)$ at various 
fields $H$.}
\label{fig:fig12}
\end{figure}

Clearly, the negative temperature coefficient of the magnetoresistance 
below 60K must be related to the GdCo/Co interface.  Note that 
this "positive" magnetoresistance (the resistance increases with the 
field) corresponds to the so-called 
"negative" magnetoresistance of the literature about DWS,
 because in contrast to our system, domain-walls 
usually disappear on increasing the external field.

\section{Analysis and discussion}

Is it possible to interpret the negative slope of the 
magnetoresistance $\Delta ^HR(T)$ (Fig.12) at low temperature and high field as 
the increasing contribution of the DWS produced by the compression of 
the domain-wall?  As the MR of the amorphous GdCo$_{1.6}$ layer is rather 
exotic and poorly understood, the effect may also be the 
manifestation of the structural disorder produced by the interface, 
and not by the domain-wall itself.

Let us first define the supplementary contribution to the magnetoresistance 
$\partial R(H,T)$. The reference point is taken with respect to the 
minimum of the $\Delta ^HR(T)$ curve at $60 \, K$ and $3.2 \, T$ (see 
Fig. 12) :

\begin{equation}
\partial R(H,T) = \Delta ^HR(T) - \Delta ^{3.2}R(60 \, K)
\label{equ2}
\end{equation}

\noindent 

Note that the reference point is chosen with respect to the 
minimum of the  $\Delta ^HR(T)$ curve at 60 K and 3.2 T which are expected to be 
the temperature and the field limits for the DW contribution (see 
paragraphs II(C) and IV).  Equation (2) describes the MR of the 
GdCo/Co interface if we assume that the Co magnetoresistance $\Delta ^HR(T)$  is 
constant below 60K (see Fig.  9). The magnetoresistance is then
expressed in terms of the domain-wall size $\delta (H,T)$, with the 
following relation \cite{34} :

\begin{equation}
\delta (H,T) = \pi \sqrt{\frac{(M_s^{Co}+M_s^{GdCo_{1.6}}(T))A}{2 \, 
M_s^{Co} \cdot M_s^{GdCo_{1.6}}(T) \cdot H}}
\label{equ3}
\end{equation}

\noindent where $M_s^{Co}$ is the saturation magnetization of Co 
(assumed constant in this temperature range) , $M_s^{GdCo_{1.6}}(T)$ 
the temperature dependent saturation magnetization of GdCo$_{1.6}$ 
deduced from the Fig3, A 
the exchange constant of Co, and H the applied magnetic field.  This 
relation was obtained by minimizing the Zeeman and domain wall 
energies.  The anisotropy contribution is neglected (approximation 
valid at $H > 3 \, T$), and the domain-wall is assumed to be centered 
at the interface.  The magnetization $M_s^{GdCo_{1.6}}(T)$ is given by 
the curve of Fig.3.  From the values of the alloy density ($\rho = 7 
\, g/cm^3$ \cite{27}) and the alloy momentum ($\mu _{GdCo_{1.6}}(2 \, 
K) = 5 \, \mu _B$ \cite{27}), we obtain $M_s^{GdCo_{1.6}}(2 \, K) = 
775 \, emu/cm^3$. 

The scaling plot $\partial R(\delta (H,T))$ is presented in Fig.13 
without any adjustable parameter (we took $A = 0.85 \cdot 10^{-6} \, 
erg/cm$ \cite{35} and $M_s^{Co} = 1430 \, emu/cm^3$ \cite{36}).  The 
scaling is rather good as the data are approximately aligned, which 
corroborates the validity of the hypothesis. The scaling plot 
corresponds to a domain wall size ranging from $\delta = 10 \, 
nm$ down to $\delta = 5 \, nm$, which is the 
range of the spin diffusion length $l_{sf}$ of Co measured in 
multilayered Cu/Co nanowires \cite{38}. The resistance of a $5 \, 
nm$ thick domain-wall is about $\partial R(4.8 \, T,2 \, K) = 0.1 
\Omega$. 

\vspace{0.7cm}
\begin{figure}[tbp]
\epsfxsize=8cm
\epsfbox{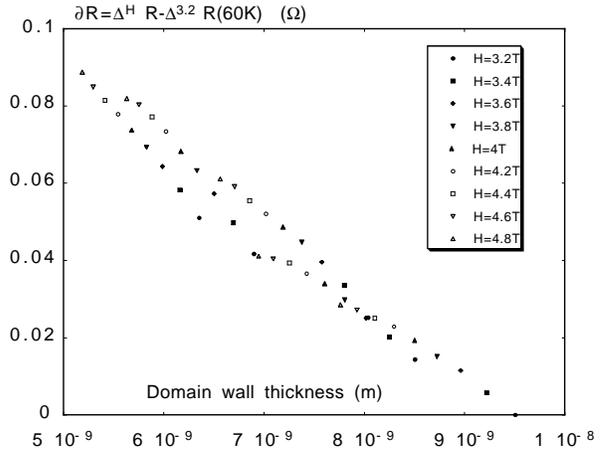}	
\caption{Scaling plot of the enhanced magnetoresistance
 $\partial R(H,T) = \Delta ^HR(T) - \Delta ^{3.2}R(60 \, K)$, as a 
function of the domain wall thickness $\delta$.}
\label{fig:fig13}
\end{figure}

 This result is in agreement with the estimation of the 
maximum interface resistance expected (i.e. in the case of infinitely 
thin domain wall) 
\cite{4,5} :

\begin{equation}
R^{GMR} = 2 \, \rho \, \beta ^2 \, l_{sf} / S
\label{equ4}
\end{equation}

\noindent where $\beta$ is the conductivity asymmetry, $l_{sf}$ is 
the spin diffusion length and S is the section of the wire ($S = 5 
\cdot10^{-15} cm^2$). With taking the values measured previously on 
multilayered nanowires obtained with the same electrodeposition 
method $\rho = 10^{-7} \, m\Omega$, $\beta = 0.45$, $l_{sf} = 20 \, 
nm$ \cite{38,6} we have $R^{GMR} = 0.2 \, \Omega$. 

Unfortunately, the small $\delta$ range in Fig13 and the rough scaling 
procedure do not allow to differentiate between the 
predictions 
$\partial R(\delta) \propto \delta ^{-2}$ (Levy and Zang in reference
\cite{8}) or $\partial R(\delta) \propto \delta ^{-1}$ and  
$\partial R(\delta) \propto -\delta $
(van Hoof et al. in reference \cite{10}).

\section{Conclusion}

Magnetic and transport properties
 have been measured on Co nanowires, on GdCo$_{1.6}$ 
thin films and on exchange coupled systems GdCo$_{1.6}$/Co.
 The ferromagnetic Co-Co exchange coupling at a GdCo$_{1.6}$/Co interface 
 induces the existence of domain-wall the 
 thickness of which can be controlled with both the external magnetic 
 field and the temperature.  An additional contribution to the 
 magnetoresistance due to the presence of this domain wall has been 
 clearly evidenced.  
 This domain wall magnetoresistance increases both with increasing the 
 magnetic field strength, and with decreasing temperature 
(the latter below 60 K only).  

A simple scaling procedure leads one to express the domain wall 
magnetoresistance as a function of the domain-wall thickness.  The 
reasonable accuracy of the scaling shows that the magnetoresistance 
enhancement is indeed due to the compression of the DW by the external 
field and/or temperature.

 The maximum domain wall contribution to the 
resistance of the GdCo$_{1.6}$/Co system is of the order of $0.1 
\Omega$ ($0.5 \, \%$ of the wire), in accordance with the GMR 
contribution of an ideal antiparallel ferromagnetic interface.

\acknowledgments

We are grateful to H. Kind (IPE,EPFL) for helping to prepare 
potassium cyanide bath. We thank U. R\"udiger and G. E. W. Bauer for helpful 
discussions.

\end{multicols}


\begin{references}

\bibitem[a)] {email} e-mail:jean-eric.wegrowe@epfl.ch
\bibitem{1} M. Johnson and R. H. Silsbee, Phys. Rev. Lett. {\bf 55}, 
1790 (1985) and M. Johnson and R.H. Silsbee Phys. Rev. B {\bf 37}, 
4959 (1987).
\bibitem{2} M. N. Baibich, J. M. Broto, A. Fert, F. Nguyen Van Dau, 
F. Petroff, Phys. Rev. Lett. {\bf 61}, 2472 (1988).
\bibitem{3} G. Binash, P. Gr\"unberg, F. Saurenbach, W. Zinn, Phys. 
Rev. B {\bf 39}, 4828 (1989).
\bibitem{39} J. S. Moodera, L.R. Kinder, T. M. Kinder T. M. Wing, R. Maservey, Phys. 
Rev. Lett. {\bf 74}, 3273 (1995).
\bibitem{4} P. C. van Son, H. van Kampen, P. Wyder, Phys. Rev. Lett. 
{\bf 58}, 7113 (1987).
\bibitem{5} T. Valet, A. Fert, Phys. Rev. B {\bf 48}, 7099 (1993).
\bibitem{38} L. Piraux, S. Dubois, C. Marchal, J.M. Beuken, L. 
Filipozzi, J.F. Despres, K. Ounadjela, A. Fert, J. Magn. Magn. Mat. {\bf 
156}, 317 (1996). 
\bibitem{6} B. Doudin, A. Blondel, J.-Ph. Ansermet, J. Appl. Phys. {\bf 
79}, 6090 (1996) and B. Voegeli, A. Blondel, B. Doudin, J.-Ph. Ansermet, J. 
Magn. Magn. Mat. {\bf 151}, 388 (1995).
\bibitem{7} J. F. Gregg, W. Allen, K. Ounadjela, M. Viret, M. Hehn, 
S.M. Thompson, J. M. D. Coey, Phys. Rev. Lett. {\bf 77}, 1580 (1996).
\bibitem{8} P. Levy and Shufeng Zhang, Phys. Rev. Lett. {\bf 79}, 
5111 (1997).
\bibitem{9} M. A. M. Gijs, G. E. W. Bauer, Adv. Phys. {\bf 46}, 285 
(1997), J.-Ph. Ansermet, J. Phys. Cond. Mat. {\bf 10}, 6027 (1998) 
and references therein.
\bibitem{10} J. B. A. N. van Hoof, K. M. Schep, A. Brataas, P. J. 
Kelly, G. E. W. Bauer, Phys. Rev. B {\bf 59}, 138 (1999).
\bibitem{11} G. Tatara and H. Fukuyama, Phys. Rev. Lett. {\bf 78}, 
3773 (1997).
\bibitem{12} A. Brataas, G. Tatara, G. E. W. Bauer, Phys. Rev. B {\bf 
60},3406 (1999) and  A. Brataas, G. Tatara, G. E. W. Bauer, Phylos. Mag. 
B {\bf 78}, 545 (1998).
\bibitem{14} U. R\"udiger, J. Yu, L. Thomas, S. S. P. Parkin, A. D. 
Kent, Phys. Rev. B {\bf 59}, 11914(1999) and U. R\"udiger, J. Yu, 
A. D. Kent, S. S. P. Parkin, J. Appl. Phys. {\bf 73}, 1298 (1998).
\bibitem{15} T. Taniyama, I. Nakatani, T. Namikawa, Y. Yamazaki, 
Phys. Rev. Lett. {\bf 82}, 2780 (1999).
\bibitem{16} K. Mibu, T. Nagahama, T. Shinjo, T. Ono, Phys. Rev. B 
{\bf 58}, 6442 (1998).
\bibitem{17} Kimin Hong, N. Giordano, J. Phys. Cond. Mat. {\bf 10}, 
L401 (1998).
\bibitem{18} U. R\"udiger, J. Yu., S. Zhang, A. D. Kent, S. S. P. 
Parkin, Phys. Rev. Lett. {\bf 80} 1116 (1998).
\bibitem{37} S.G. Kim, Y. Otani, K. Fukamichi, S. Yuasa, M. Nyvlt, T. 
Katayama,  J. Magn. Magn. Mat. {\bf 198-199}, 
200 (1999).
\bibitem{19} T. R. McGuire and R. I. Potter, IEEE Trans. Magn. {\bf 
MAG-11}, 1018 (1975).
\bibitem{20} Th. G. S. M. Rijks, R. Coehoorn, M. J. M. de Jong, W. J. 
M. de Jonge, Phys. Rev. B {\bf 51}, 283 (1995).
\bibitem{23} J.-E. Wegrowe, S. E. Gilbert, V. Scarani, D. Kelly, B. 
Doudin, J.-Ph. Ansermet, IEEE Trans. Magn. {\bf 34}, 903 (1998).
\bibitem{21} S. W\"uchner, J. Voiron, D. Givord, D. Boursier, J. J. 
Pr\'ejean, J. Appl. Phys. {\bf 75}, 6682 (1994).
\bibitem{22} J.-E. Wegrowe, B. Barbara, V. S. Amaral, J. B. Sousa, J. 
Magn. Magn. Mat. {\bf 161}, 133 (1996).
\bibitem{24} I. Chlebny, B. Doudin, J.-Ph. Ansermet, Nanostructured 
Materials {\bf 2}, 637 (1993).
\bibitem{25} V. Scarani, B. Doudin, J.-Ph. Ansermet, to appear in J. 
Magn. Magn. Mat. (1999).
\bibitem{26} J.-E. Wegrowe, D. Kelly, A. Franck, S. E. Gilbert, 
J.-Ph. Ansermet, Phys. Rev. Lett. {\bf 82}(18), 3681 (1999).
\bibitem{27} R. Ferr\'e, K. Ounadjela, J. M. George, L. Piraux and S. 
Dubois, Phys. Rev. B {\bf 56}, 14066 (1997).
\bibitem{28} S. W\"uchner, doctoral thesis, Universit\'e Joseph 
Fourier, Grenoble France (1995).
\bibitem{29} A. Fert and D.K. Lottis in "Concise encyclopedia of 
magnetic and superconducting materials" Ian Evettts editor, Pergamon 
Press, p. 291
\bibitem{30} R. W. Cochrane, R. Harris, J. O. Str–m-Olson, M. 
Zuckermann, Phys. Rev. Lett. {\bf 35}, 676 (1995). 
\bibitem{31} D. L. Cox, A. Zawadowshi, Adv. Phys. {\bf 47}, 599-942 
(1998).
\bibitem{32} H. Okuno, Y. Sakurai, J. Magn. Magn. Mat. {\bf 35}, 80 
(1983), H. Okuno, Y. Sakurai, J. Appl. Phys. {\bf 53}, 8245 (1982), 
H. Okuno, Y. Sakurai, IEEE Trans. Magn. {\bf 17}, 2831 (1981).
\bibitem{34} B. Dieny, D. Givord, J. M. B. Ndjaka, J. Magn. Magn. 
Mat. {\bf 93}, 503 (1991).
\bibitem{35} P. Gaunt and C. K. Mylvaganam, Phil. Mag. B {\bf 44}(5), 
569 (1981).
\bibitem{36} E. K\"oster in "magnetic recording technology" (C.D. Mee 
and E.D. Daniel. eds.) chapter 3, McGraw-Hill, New-York (1995).


\end{references}
\end{document}